# Significant enhancement of flux pinning in $MgB_2$ superconductor through nano-Si addition

**X.L. Wang[1], S.H. Zhou[1], M.J. Qin[1], P. R. Munroe[2], S. Soltanian[1], H.K. Liu[1], and S.X. Dou[1]**
*[1]Institute for Superconducting and Electronic Materials, University of Wollongong, North Ave. NSW 2552, Australia*
*[2] Electron Microscope Unit, University of New South Wales, NSW 2052, Australia*

Polycrystalline $MgB_2$ samples with 10 wt % silicon powder addition were prepared by an in-situ reaction process. Two different Si powders, one with coarse (44 μm) and the other with nano-size (<100 nm) particles were used for making samples. The phases, microstructures, and flux pinning were characterized by XRD, TEM, and magnetic measurements. It was observed that the samples doped with nano-sized Si powder showed a significantly improved field dependence of the critical current over a wide temperature range compared with both undoped samples and samples with coarse Si added. Jc is as high as 3000 A/cm$^2$ in 8 T at 5 K, one order of magnitude higher than for the undoped $MgB_2$. X-ray diffraction results indicated that Si had reacted with Mg to form $Mg_2Si$. Nano-particle inclusions and substitution, both observed by transmission electron microscopy, are proposed to be responsible for the enhancement of flux pinning in high fields. However, the samples made with the coarse Si powders had a poorer pinning than the undoped $MgB_2$.

PACS: 74.70. Ad, 74.70.Ge, 74.62 BF, 74.70. Jg

To push the newly discovered $MgB_2$ superconductor towards practical applications, it is essential to enhance the flux pinning and increase the critical current density in high fields to the level attained by the conventional low temperature superconductors NbTi or $Nb_3Sn$. Among all the approaches aimed at enhancement of flux pinning in $MgB_2$ such as high energy ion irradiation [1,2] for powders of $MgB_2$, inclusion of precipitates of MgO in thin films [3], and chemical doping using different elements, nano-particle addition/doping seems to be the best and most practical route [3, 4] reported so far. Very recently, Dou et al [5] have demonstrated that nano-size SiC-doped $MgB_2$ samples show a very significant enhancement of critical current density in high magnetic fields over a wide temperature range from 30 K down to 5 K. A newly reported of $J_c$ as high as 20,000 A/cm$^2$ in 9.5 T at 5 K was achieved for both transport and magnetic measurements [6,7]. $Mg_2Si$ phase was found to be the major impurity phase in the SiC-doped $MgB_2$ samples [6]. $Mg_2Si$ inclusions inside $MgB_2$ grains are suggested to be one of the possible sources responsible for the enhanced pinning in addition to SiC substitutions. The intention of the present work is to study the effect of pure nano-size Si addition on the phase formation and microstructures of $MgB_2$ and to investigate its role in the flux pinning behaviors. Our present results indicate that, quite similar to nano-SiC doped $MgB_2$ [5-7], pure nano-Si doped $MgB_2$ exhibits significantly enhanced pinning in high fields over a wide temperature range compared with undoped $MgB_2$ samples. Both Si substitution and nano-particle inclusions were observed by transmission electron microscopy, leading to a significant enhancement in flux pinning.

It has been reported that $MgB_2$ with high performance can be formed very quickly (in as little as a few minutes) [8] in a reaction between individual magnesium and amorphous boron powders, the so-called "in-situ" reaction. It has also been demonstrated that the $J_c$ magnitude and its field dependence in low density $MgB_2$/Fe wire [7] fabricated using the "in-situ" is better than that those of fully dense $MgB_2$/Fe tapes [9] made using pre-reacted $MgB_2$ powders (ex-situ reaction). The samples used in the present study were thus made using the "in-situ" solid state reaction. Magnesium powders (-325 mesh) and amorphous boron (-325 mesh) powders were well mixed with silicon powders with different particle sizes, 44 μm (coarse) and <100 nm (fine) at a level of 10 wt%. Pellets 10 mm in diameter and 1 mm in thickness were made under pressure, and sealed in Fe tubes, then heat treated at 700-900 C for 1 hour in flowing high purity Ar. This was followed by a furnace cooling to room temperature. An undoped sample was also made under the same conditions for use as a reference sample. The phases and microstructures were examined by X-ray diffraction (XRD) and transmission electron microscopy (TEM). The magnetization was measured over a wide temperature range between 5 and 30 K in 0-8.5 T using a Quantum Design PPMS. Bar shaped samples with dimensions of a x b x c=0.56x2.7x3.73

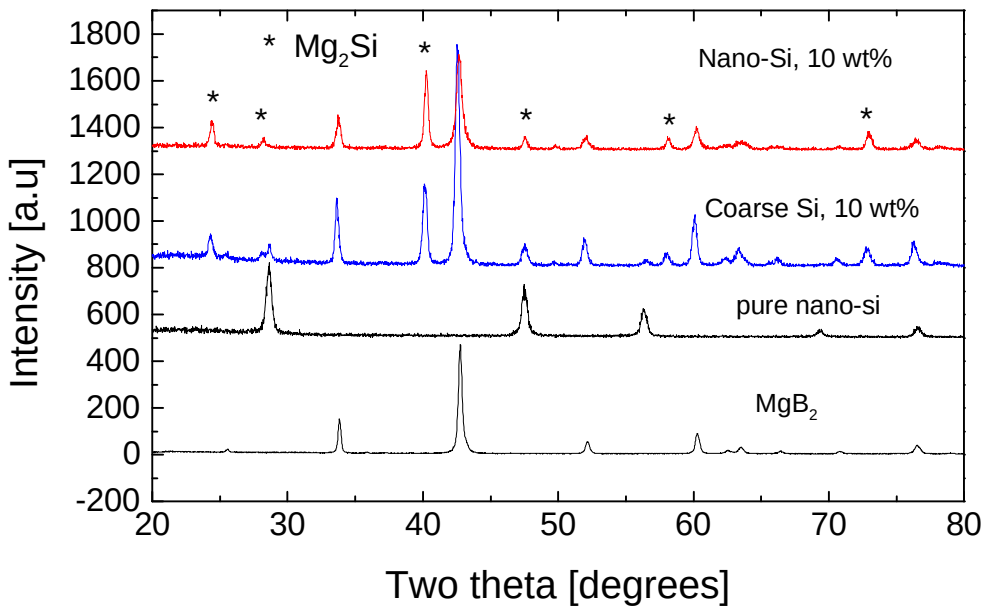

Fig. 1. XRD pattern for undoped, coarse-Si and nano-Si doped MgB2, and pure nano-Si powder samples.

mm$^3$ were cut from as-sintered pellets. Transition temperatures were determined by measuring the real part of the ac susceptibility at a frequency of 117 Hz and an ac field of 0.1 Oe. The magnetic $J_c$ was calculated from the height of the magnetization loops (M-H) using a Bean model where $J_c$=20 $\Delta$M/[a/(1-a/3b)] with a<b, where a and b are the dimensions of the sample perpendicular to the direction of applied field.

Fig.1. shows the XRD patterns for four samples. These are reference sample (sample #1), coarse-Si (sample 2#) and nano-Si (sample #3) doped samples, and the pure nano-si powder, respectively. It can be clearly seen that $Mg_2Si$ is the only major impurity phase in both the coarse and nano-Si doped samples. (A very small amount of MgO can also be observed from the peak located at 2$\Theta$ =62.4 degrees). Some Si peaks can be found in the coarse-Si doped sample in addition to the $Mg_2Si$, indicating that there is residual unreacted powder. However, the nano-Si powder seems to react completely with Mg and form $Mg_2Si$ or substitute into the $MgB_2$ lattice due to the higher reactivity of nano-Si compared to coarse Si powders. This is in agreement with the fact that the relative XRD peak intensity from the $Mg_2Si$ phase to that from $MgB_2$ in the nano-Si doped sample is higher than that in the coarse-Si doped one.

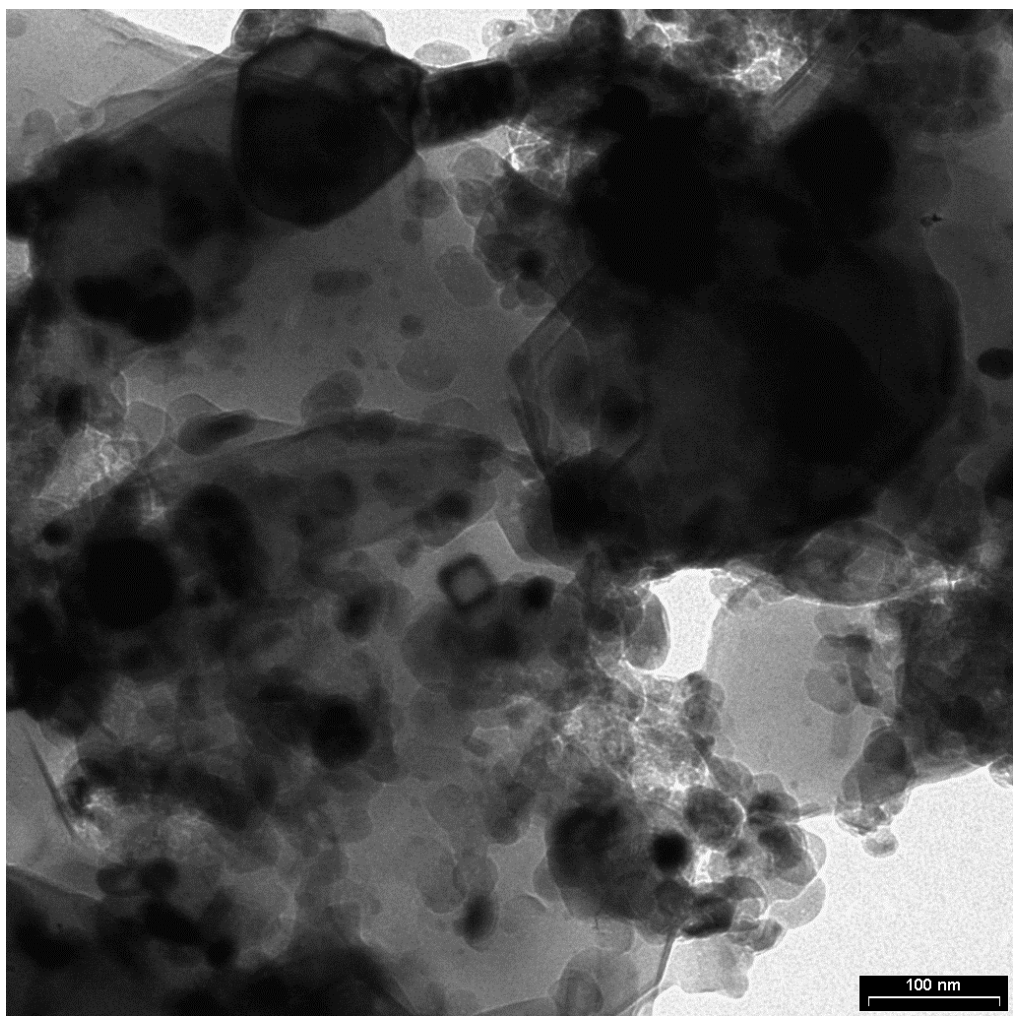

Fig. 2. TEM micrograph of nano-Si doped $MgB_2$.

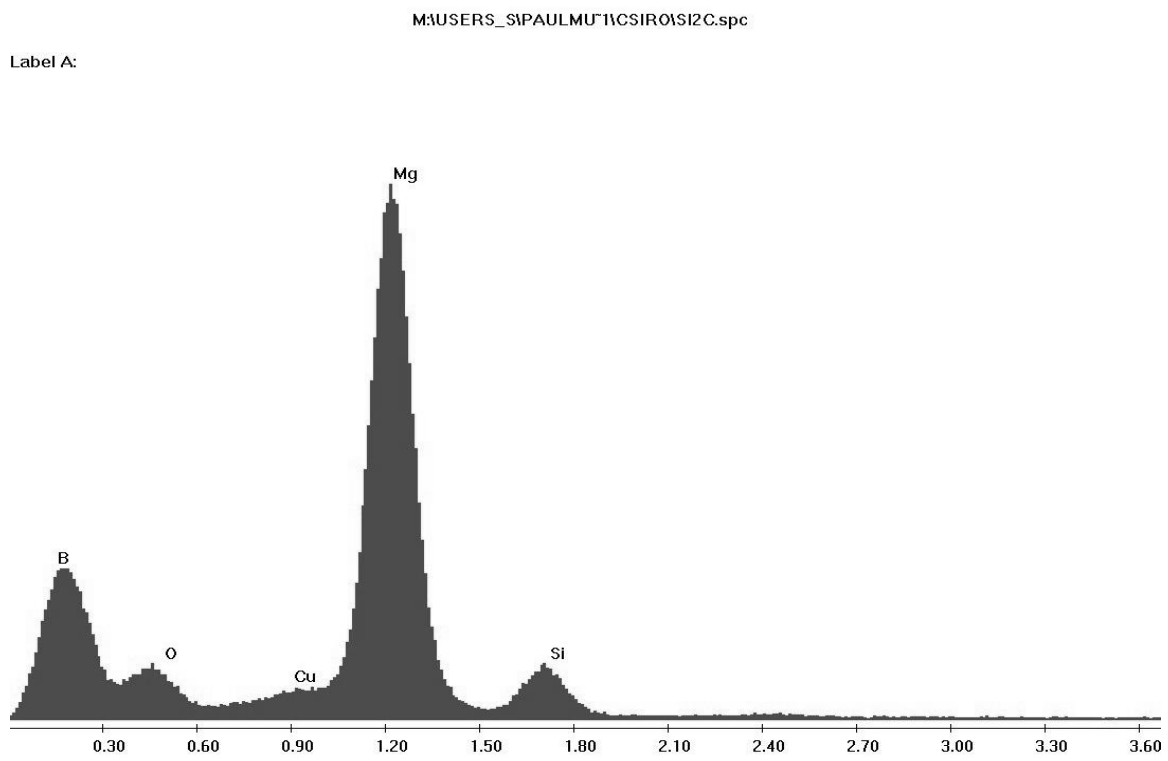

Fig. 3. Typical EDS spectrum taken from one of the grains shown in Fig. 2.

A typical TEM image of $MgB_2$ grains from a nano-Si doped $MgB_2$ sample is shown in Fig. 2. A number of fine grains, around 100-200nm in diameter can be seen. Most of these were shown by electron diffraction to be $MgB_2$, but a number of grains, especially those somewhat larger than 100nm in diameter, were determined by electron diffraction to be $Mg_2Si$. EDS analysis was carried out on several spots in different regions of the same $MgB_2$ grain. A small Si peak was always detected, but varied slightly in intensity at different positions. A typical EDS spectrum is shown in Fig. 3. EDS spectra of the $Mg_2Si$ grains yielded significantly more intense Si peaks. A number of smaller particles, 10-30nm in size were also observed located on the $MgB_2$ or $Mg_2Si$ grains, but EDS and diffraction data from these particles were not conclusive as the signal from these particles is diluted by the surrounding grains.

There are two possibilities for the form of the Si detected in the $MgB_2$ grains: first, that it is in the form of either $Mg_2Si$ or Si inclusions with very fine particle sizes, or second, that it is Si which has substituted directly into the $MgB_2$ crystal lattice. The former may be possible as $Mg_2Si$ was the only impurity detected by XRD. EDS and microdiffraction studies are potentially ambiguous since the typical probe size used (~10nm) may be larger than the size of the $Mg_2Si$ or Si particles, and such particles may be embedded in thicker $MgB_2$ crystals, where the signal from the Si-rich regions is diluted by the surrounding $MgB_2$. However, no Si was observed via electron diffraction, hence we suggest that the Si peak in EDS spectra may partly come from Si which occupies substitutional positions in the $MgB_2$ and partly from $Mg_2Si$ inclusions inside the $MgB_2$ grains.

The real part of the ac susceptibility of all three samples is shown in Fig. 4. We can see that the $T_c$ of the coarse-Si doped sample is slightly higher than that of the undoped sample despite the large amount of $Mg_2Si$ and the presence of residual Si. However, the $T_c$ of the nano-Si doped sample has dropped to 36.7 K which is higher than that of the 10 wt % nano-SiC doped $MgB_2$. The drop in the $T_c$ for the nano-Si doped sample is another piece of evidence for substitution of Si into the lattice. For the coarse Si doping, the level of substitution into the lattice is very small due to the limited contact area between the Si and the Mg and B.

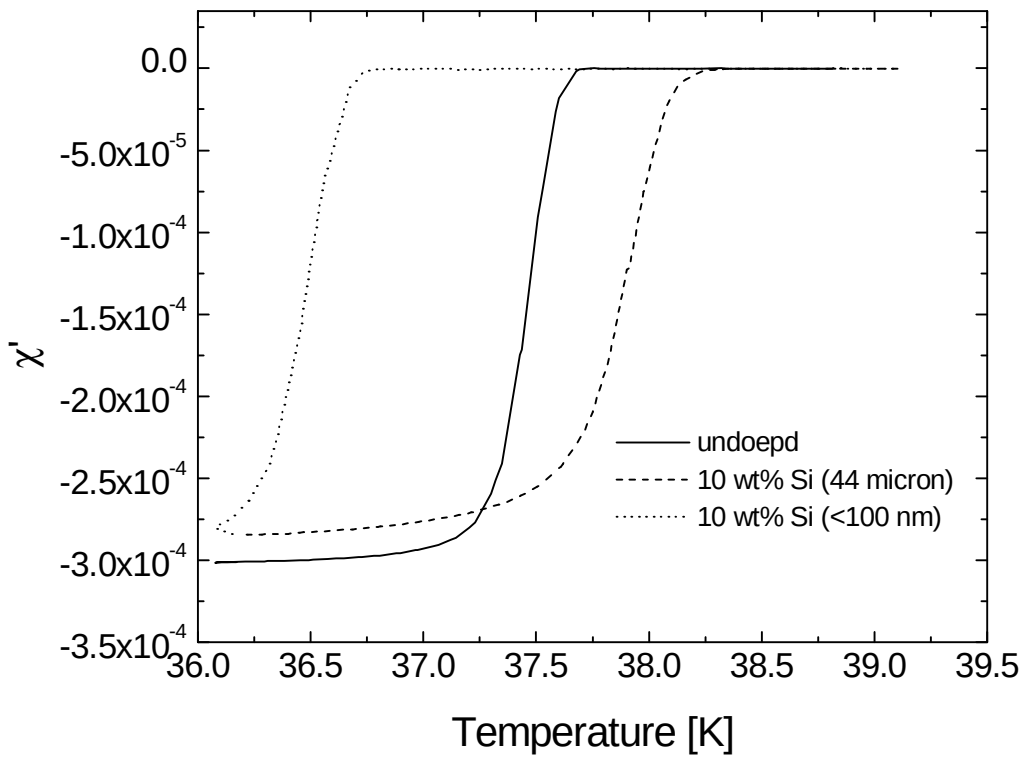

Fig.4. Real part of the ac susceptibility for coarse- Si doped, nano-Si doped and undoped $MgB_2$ samples.

whereas the nano-Si has much high surface energy and can react with Mg and B more fully. Thus the substitution level is expected to be much higher than is the case for coarse Si.

Values for $J_c$ vs field at 30, 20, 10, and 5 K are plotted in Fig. 5. The coarse Si-doped sample shows a poor $J_c$ field dependence at T< 30 K, compared to the undoped sample. However, the nano-Si doped sample revealed a significant improvement in $J_c$ field performance at all temperatures. The irreversibility field $H_{irr}$ and $J_c$ values of the nano-Si doped sample are higher than for the coarse-Si doped and undoped samples in high field regions at all the measured temperatures. The lower the temperature, the larger the difference in $H_{irr}$ and $J_c$ at high fields for these three samples. It can be seen that the $J_c$ for the nano-Si doped sample drops more slowly than for the reference sample as the field increases. However, $J_c$ values for coarse-Si doped sample are much smaller than for the undoped sample and decrease in the same way, indicating that there is no improved pinning as a result of the coarse-Si doping. No nano-inclusion features were found in samples doped with coarse-Si powders, rather large grains of Si and $Mg_2Si$ were observed around $MgB_2$ particles in SEM images, and this would create harmful weak links between the $MgB_2$ grains.

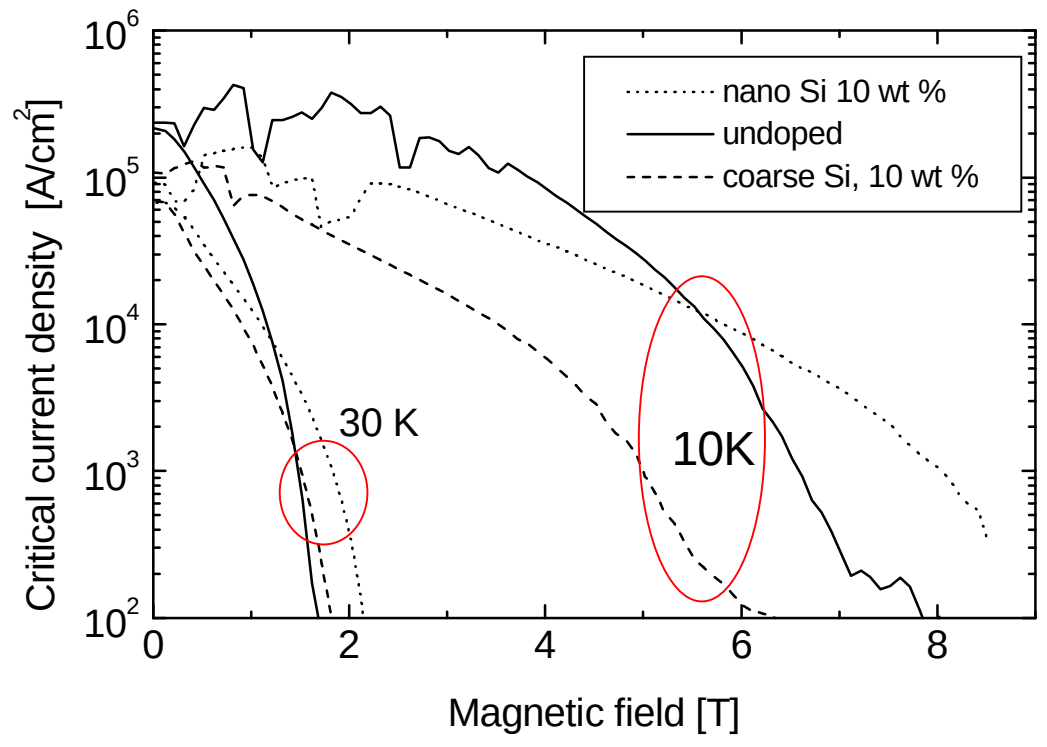

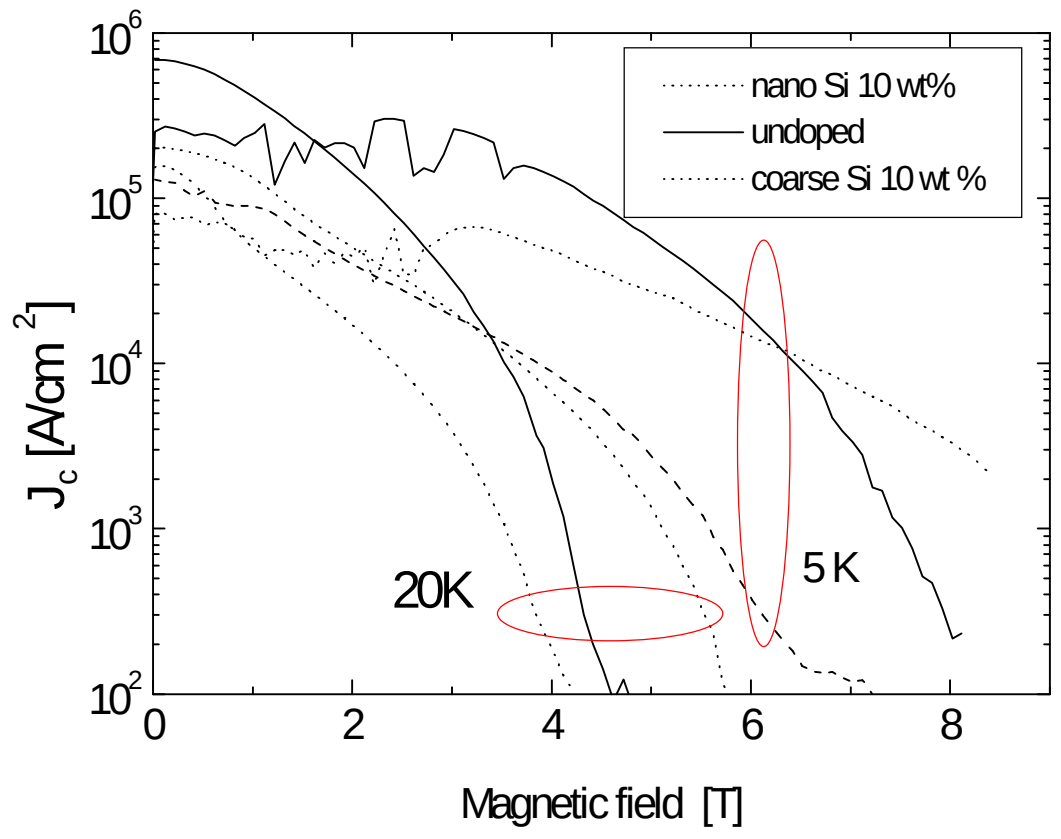

Fig. 5. Jc vs field for undoped, coarse-Si and nano-Si doped samples at 30 and 10 K (upper) and at 20 and 5 K (lower).

A $J_c$ of 3000 A/cm2 is achieved in 8 T at 5 K for the nano-Si doped $MgB_2$, one order of magnitude higher than for the control sample. This Jc field behavior in high fields is similar to that observed in nano-SiC doped $MgB_2$ [6], but it is better than for nano-$Y_2O_3$ doped $MgB_2$ [4], the best pure $MgB_2$ samples [10] which were sintered in Mg vapor, or fully dense $MgB_2$ made using the hot pressure method [11]. The enhancement of flux pinning in high fields may be ascribed to both substitution effects and fine nano-particle inclusions inside $MgB_2$ grains.

It should be noted that there is a clear difference between nano-additive induced pinning such as in the case of nano-$Y_2O_3$, and the situation for nano-SiC and nano-Si. The former gives a strong enhancement of flux pinning at low temperatures but not at higher temperatures (above 20K). In contrast, the nano-SiC and nano-Si doping with substitution as well as induced nano-particle inclusions exhibit strong enhancement in flux pinning over all the entire temperature range despite the fact that the $T_c$ was depressed more than for the nano-particle addition. Thus, it is essential to introduce intra-grain pinning into $MgB_2$ to achieve the best high field performance..

Only a 10 wt% addition of Si was added to the $MgB_2$ in our present work. It is believed that a further enhancement can be achieved when the nano-Si addition level and the heat treatment conditions are further optimized.

In summary, nano-Si particle sizes (less than 100 nm) doped $MgB_2$ demonstrated significantly improved pinning in high fields over a wide temperature range compared to undoped $MgB_2$. In contrast, a coarse (44 um) Si doped sample had weaker pinning than pure $MgB_2$. TEM results indicated that both Si substitional effects and fine particle inclusions inside $MgB_2$ grains are responsible for the improved intra-grain flux pinning in the nano-Si doped samples.

Acknowledgements

The authors thank Drs. T. Silver, M. Ionescu, J. Horvat, E.W. Collings, M. Sumption, M. Tomsic and R. Neale for their helpful discussion and Australian Research Council, Hyper Tech Research Inc. OH USA and Alphatech International Ltd for financial support.